\documentclass[preprint,showpacs,preprintnumbers,amsmath,amssymb]{revtex4}
\usepackage{epsfig}
\usepackage{graphicx}
\usepackage{dcolumn}
\usepackage{bm}
\usepackage{threeparttable}

\def\beq{\begin{equation}}
\def\eeq{\end{equation}}
\def\eeqn{\end{equation}}
\newcommand\iden{\leavevmode\hbox{\small1\normalsize\kern-.33em1}}

\newcommand{\bea} {\begin{eqnarray}}
\newcommand{\eea} {\end{eqnarray}}

\let\jnfont=\rm
\def\NPB#1,{{\jnfont Nucl.\ Phys.\ B }{\bf #1},}
\def\PLB#1,{{\jnfont Phys.\ Lett.\ B }{\bf #1},}
\def\EPJC#1,{{\jnfont Eur.\ Phys.\ Jour.\ C }{\bf #1},}
\def\PRD#1,{{\jnfont Phys.\ Rev.\ D }{\bf #1},}
\def\PRL#1,{{\jnfont Phys.\ Rev.\ Lett.\ }{\bf #1},}
\def\MPLA#1,{{\jnfont Mod.\ Phys.\ Lett.\ A }{\bf #1},}
\def\JPG#1,{{\jnfont J.\ Phys.\ G }{\bf #1},}
\def\CTP#1,{{\jnfont Commun.\ Theor.\ Phys.\ }{\bf #1},}
\def\JHEP#1,{{\jnfont JHEP \ }{\bf #1},}
\def\NPPS#1,{{\jnfont Nucl.\ Phys.\ Proc.\ Suppl.\ }{\bf #1},}
\def\CPC#1,{{\jnfont Comput.\ Phys.\ Commun.\ }{\bf #1},}
\def\CPL#1,{{\jnfont Chin.\ Phys.\ Lett. }{\bf #1},}
\def\APPB#1,{{\jnfont Acta\ Phys.\ Polon.\ B }{\bf #1},}

\def\lsim{\raise0.3ex\hbox{$<$\kern-0.75em\raise-1.1ex\hbox{$\sim$}}}
\def\gsim{\raise0.3ex\hbox{$>$\kern-0.75em\raise-1.1ex\hbox{$\sim$}}}
\def\PR#1,{{\jnfont Phys.\ Rept. }{\bf #1},}
\def\CHC#1,{{\jnfont Chin.\ Phys.\ C }{\bf #1},}
\def\ZPC#1,{{\jnfont Zeit.\ Phys.\ C }{\bf #1},}
\begin{document}

\title{\ \\[10mm] Explaining 750 GeV diphoton excess from top/bottom partner cascade decay in two-Higgs-doublet model extension}
\author{Xiao-Fang Han$^{1}$, Lei Wang$^{2,1}$, Lei Wu$^{3}$, Jin Min Yang$^{4,5}$, Mengchao Zhang$^{4}$}
 \affiliation{$^1$ Department of Physics, Yantai University, Yantai
264005, P. R. China\\
$^2$ IFIC, Universitat de Val$\grave{e}$ncia-CSIC, Apt. Correus
22085, E-46071 Val$\grave{e}$ncia, Spain\\
$^3$ ARC Centre of Excellence for Particle Physics at the Terascale,
School of Physics, The University of Sydney, NSW 2006, Australia
\\
$^4$ Institute of Theoretical Physics, Academia Sinica, Beijing 100190, China\\
$^5$ Department of Physics, Tohoku University, Sendai 980-8578, Japan
}


\begin{abstract}
In this paper, we interpret the 750 GeV diphoton excess in the Zee-Babu extension of the
two-Higgs-doublet model by introducing a top partner ($T$)/bottom partner ($B$).
In the alignment limit, the 750 GeV resonance is identified as the heavy CP-even Higgs boson ($H$), which can be sizably
produced via the QCD process $pp \to T\bar{T}$ or $pp \to B\bar{B}$ followed by the
decay $T\to Ht$ or $B \to Hb$.
The diphoton decay rate of $H$ is greatly enhanced by the charged singlet scalars
predicted in the Zee-Babu extension and the total width of $H$ can be as large as
7 GeV. Under the current LHC constraints, we scan the parameter space and find that
such an extension can account for the observed diphoton excess.
\end{abstract}
 \pacs{12.60.Fr, 14.80.Ec, 14.80.Bn}

\maketitle

\section{Introduction}
Very recently, both the ATLAS data with 3.2 fb$^{-1}$ and the CMS
data with 2.6 fb$^{-1}$ \cite{750} have reported an excess of the
diphoton resonance ($X$) around 750 GeV. The local significances of
their results are $3.6\sigma$ and $2.6\sigma$ in the respective
experiments. Combining the 8 and 13 TeV data \cite{1512.04939}, the
observed signal strength $\sigma_X \times Br(X \to \gamma\gamma)$ is
$10.6 \pm 2.9$ fb for the ATLAS and $4.47 \pm 1.86$ fb for the CMS.
Since there are no excesses observed in the dijet \cite{dijet},
$t\bar{t}$ \cite{ditt}, diboson or dilepton channels, understanding
such an excess becomes a challenging task. So far, many new physics
models have been proposed for this excess
\cite{1512.04939,work1,work2,work3,work4,work5,work6,work7,work8},
among which, a singlet scalar is usually introduced as the 750 GeV
resonance.

Different from the previous singlet scalar explanations, we attempt to interpret the 750 GeV resonance
as a heavy Higgs boson from a second doublet, which
is mainly originating from the QCD top partner ($T$) or bottom partner ($B$) pair production
process followed by the decay $T\to Ht$ or $B \to Hb$. Obviously, such a scenario still needs the extra
particles to enhance the 750 GeV Higgs decay into diphoton. Therefore, we introduce a top partner/bottom partner
to the Zee-Babu extension \cite{zeebabu} of the two-Higgs-doublet
model (ZB-2HDM), where two extra charged singlet scalars can enhance the decay of diphoton mode and
generate the neutrino mass. Considering the LHC Higgs data, our study will be focused on an interesting limit of this model,
in which one of the neutral Higgs mass eigenstates is almost aligned with the direction of the
scalar field vacuum expectation values. In this limit, the 125 GeV Higgs boson tends to have the
gauge couplings as in the Standard Model (SM) and is easily consistent with the current Higgs
data, while the heavy CP-even Higgs boson has the very small couplings or no couplings to the SM particles.

Compared to the direct $gg \to H$ production process, there are several benefits for the production of $H$ from
 the QCD process $pp \to T\bar{T}/B\bar{B} \to HH +t\bar{t}/b\bar{b}$. Since the production of $T/B$ and the decay of $H$ are
generally unrelated, it is easy to obtain a large branching ratio of $H \to \gamma\gamma$ by suppressing
the 750 GeV Higgs coupling to the top quark.
Although the cascade decays have other
objects in the diphoton events, such as the additional top or bottom quark jets, the status of whether or not there are other objects in the
event is unclear at the moment. So, currently, the cascade decay is still a feasible way to
interpret the 750 GeV diphoton excess although not very likely.

Our work is organized as follows. In Sec. II we present the Zee-Babu extension of the 2HDM
with the top/bottom partner. In Sec. III we perform the numerical calculations and discuss
the 750 GeV diphoton production rate and the total width of the resonance in the allowed
parameter space. Finally, we give our conclusion in Sec. IV.

\section{Model}
\subsection{Two-Higgs-doublet model}

The general Higgs potential is written as \cite{2h-poten}
\begin{eqnarray} \label{V2HDM} \mathrm{V} &=& \mu_1^2
(\Phi_1^{\dagger} \Phi_1) + \mu_2^2 (\Phi_2^{\dagger}
\Phi_2) + \left[\mu_3^2 (\Phi_1^{\dagger} \Phi_2 + \rm h.c.)\right]\nonumber \\
&&+ \lambda_1  (\Phi_1^{\dagger} \Phi_1)^2 +
\lambda_2 (\Phi_2^{\dagger} \Phi_2)^2 + \lambda_3
(\Phi_1^{\dagger} \Phi_1)(\Phi_2^{\dagger} \Phi_2) + \lambda_4
(\Phi_1^{\dagger}
\Phi_2)(\Phi_2^{\dagger} \Phi_1) \nonumber \\
&&+ \left[\lambda_5 (\Phi_1^{\dagger} \Phi_2)^2 + \rm
h.c.\right]+ \left[\lambda_6 (\Phi_1^{\dagger} \Phi_1)
(\Phi_1^{\dagger} \Phi_2) + \rm h.c.\right] \nonumber \\
&& + \left[\lambda_7 (\Phi_2^{\dagger} \Phi_2) (\Phi_1^{\dagger}
\Phi_2) + \rm h.c.\right].
\end{eqnarray}
Here we focus on the CP-conserving case where all
$\lambda_i$ and $m_{12}^2$ are real. In the Higgs basis, the two complex
scalar doublets with the hypercharge $Y = 1$ can be written as
\begin{equation} \label{field}
\Phi_1=\left(\begin{array}{c} G^+ \\
\frac{1}{\sqrt{2}}\,(v+\rho_1+iG_0)
\end{array}\right)\,, \ \ \
\Phi_2=\left(\begin{array}{c} H^+ \\
\frac{1}{\sqrt{2}}\,(\rho_2+iA)
\end{array}\right).
\end{equation}
The $\Phi_1$ field has the vacuum expectation value (VEV) $v=$246
GeV, and the VEV of $\Phi_2$ field is zero. The $G^0$ and $G^+$ are
the Nambu-Goldstone bosons which are eaten by the gauge bosons. The
$H^+$ and $A$ are the mass eigenstates of the charged Higgs boson and
CP-odd Higgs boson, and their masses are given by \beq\label{mamhp}
m_A^2=m^2_{H^{\pm}}+ v^2(\frac{1}{2}\lambda_4-\lambda_5). \eeq

The physical CP-even Higgs bosons $h$ and $H$ are the linear combination of $\rho_1$
and $\rho_2$,
\begin{equation}\label{mixhiggs}
\left(\begin{array}{c} \rho_1 \\
\rho_2
\end{array}\right)\, =\ \
\left(\begin{array}{c} \sin\theta~~~\cos\theta \\
\cos\theta~-\sin\theta
\end{array}\right)\,
\left(\begin{array}{c} h \\
H
\end{array}\right)\,.
\end{equation}
To satisfy the 125 GeV Higgs data, we focus on the so-called
\emph{alignment limit} \cite{haber},
which corresponds to $\lambda_6 =0$ and $\cos\theta =0$. In this limit, the two CP-even Higgs masses
are given as
\beq\label{mhmhp}
m_h^2=2\lambda_1 v^2,~~~~~~m_H^2=m^2_{H^{\pm}}+ v^2(\frac{1}{2}\lambda_4+\lambda_5).
\eeq

The general Yukawa interactions without the tree-level FCNC can be
given by \cite{a2hm} \bea \label{yukawacoupling} - {\cal L} &=& y_u\,\overline{Q}_L \,
(\tilde{{ \Phi}}_1 + \kappa_u \tilde{{ \Phi}}_2) \,u_R +\,y_d\,
\overline{Q}_L\,({\Phi}_1 \,+\, \kappa_d\,
{\Phi}_2) \, d_R   \nonumber \\
&&\hspace{-3mm}+ \, y_l\,\overline{L}_L \, ({\Phi}_1
\,+\, \kappa_\ell\, {\Phi}_2)\,e_R \,+\, \mbox{h.c.}\,, \eea
where $Q_L^T=(u_L\,,d_L)$, $L_L^T=(\nu_L\,,l_L)$, and
$\widetilde\Phi_{1,2}=i\tau_2 \Phi_{1,2}^*$. $y_u$, $y_d$ and $y_\ell$ are
$3 \times 3$ matrices in family space, and $\kappa_u$, $\kappa_d$ and $\kappa_\ell$ are the coupling constants.
The couplings of neutral Higgs bosons normalized to the SM Higgs boson
are give by \bea \label{normacoupling}&&
y^h_{V}=\sin\theta,~~~y^h_f=\sin\theta+\cos\theta\kappa_f,\nonumber\\
&&y^H_{V}=\cos\theta,~~~y^H_f=\cos\theta-\sin\theta\kappa_f,\nonumber\\
&&y^A_{V}=0,~~~~~y^A_u=-i\gamma^5
\kappa_{u},~~~~~y^A_{d,\ell}=i\gamma^5 \kappa_{d,\ell},\eea
where $V$ denotes $Z$ and $W$, and $f$ denotes $u$, $d$ and $\ell$.

\subsection{Zee-Babu Extension}
In order to enhance the branching ratio of the 750 GeV Higgs boson decay to diphoton,
we can suppress the total width by taking a small
heavy CP-even Higgs coupling to the top quark. However, for this case the charged
Higgs of 2HDM ($H^{\pm}$) can not enhance the branching ratio of
diphoton sizably. The perturbativity will give the upper bound of the the heavy CP-even Higgs coupling to the charged Higgs.
 A light $H^{\pm}$ can enhance the width of
$H\to \gamma\gamma$, but the decay $H\to H^\pm W^\mp$ will be open
and enhance the total width more sizably. Therefore, some additional
particles are needed to enhance the 750 GeV Higgs decay into
diphoton, such as the vector-like fermions or the charged scalars.
Since the amplitude of $H\to \gamma\gamma$ is proportional to the
square of electric charge of the particle in the loop, the
multi-charged particle can enhance $H\to \gamma\gamma$ sizably.

Here we take the approach of Zee-Babu model to introduce two $SU(2)_L$ singlet
scalar fields $\pi^+$ and $\chi^{++}$ with hypercharge 1 and 2 \cite{zeebabu}, respectively.
In addition to enhancing the decay rate of $H\to \gamma\gamma$ sizably, this model can naturally give rise to the small neutrino Majorana mass.

The potential of the two singlet scalars can be written as
\bea\label{zeebabu}
V&=&m_\pi^2 \pi^+ \pi^- + m_\chi^2 \chi^{++}\chi^{--} + k_1 \Phi_1^{\dagger} \Phi_1 \pi^+ \pi^- + k'_1 \Phi_1^{\dagger} \Phi_1 \chi^{++}\chi^{--} \nonumber\\
&&+ k_2 \Phi_2^{\dagger} \Phi_2 \pi^+ \pi^- + k'_2 \Phi_2^{\dagger} \Phi_2 \chi^{++}\chi^{--}
+ k_3 (\Phi_1^{\dagger} \Phi_2 + \Phi_2^{\dagger} \Phi_1)\pi^+\pi^- \nonumber\\
&&
+ k_4 (\Phi_1^{\dagger} \Phi_2 + \Phi_2^{\dagger} \Phi_1)\chi^{++}\chi^{--}
+ k_5 (\pi^+ \pi^-)^2 + k_6 (\chi^{++} \chi^{--})^2 + (\mu
\pi^-\pi^-\chi^{++} + h.c.). \eea The gauge invariance precludes the
singlet Higgs fields from coupling to the quarks. The Yukawa
coupling of singlets to leptons are \beq
 {\cal L} =f_{ab}\bar{L_{La}^C}L_{Lb}\pi^+ + g_{ab}\bar{E_{Ra}^C}E_{Rb}\chi^{++} + h.c..
\eeq The trilinear $\mu$ term in Eq. (\ref{zeebabu}) breaks the
lepton number and gives rise to the neutrino Majorana mass
contributions at the two-loop level. The detailed introductions on
the neutrino mass can be found in \cite{zeebabu}. Here we focus on the charged
Higgs couplings to the heavy CP-even Higgs. Since the $k_1$ and
$k_1'$ terms of Eq. (\ref{zeebabu}) that contain the 125 GeV Higgs
couplings to charged Higgs are proportional to $\sin\theta$,
we assume $k_1$ and $k_1'$ to be very small and ignore them in our calculations.
Then, after the $\Phi_1$ acquires the VEV, the masses of
$\pi^+$ and $\chi^{++}$ are $m_\pi$ and $m_\chi$, and the CP-even
Higgs couplings to the charged Higgses are determined by $k_3$ and $k_4$ terms,
\bea
&&h\pi^+\pi^-:-k_3\cos\theta v,~~~~h\chi^{++}\chi^{--}:-k_4\cos\theta v,\nonumber\\
&&H\pi^+\pi^-:k_3\sin\theta v,~~~~~H\chi^{++}\chi^{--}:k_4\sin\theta v.
\eea

For $\cos\theta=0$, the couplings of $h\pi^+\pi^-$ and $h\chi^{++}\chi^{--}$ are zero.
Considering the constraints of perturbativity and stability of the potential, we simply take $0\lesssim k_3=k_4 \lesssim 4\pi$, and fix $m_\pi=m_\chi=375$ GeV,
which will give the maximal value of the form factor of scalar loop in the $H\to \gamma\gamma$ decay.

\subsection{Top/Bottom Partners}
Next, we introduce the top partner to interact with $\Phi_2$ in the 2HDM. The Yukawa interaction is given as \beq - {\cal L} = y_T\,
\bar{Q}_{tL} \, \widetilde\Phi_2 T_R  + m_T\bar{T}_LT_R +
m'_T\bar{T}_Lt_R + h.c.,
 \label{partner}\eeq
where $Q^{T}_{tL}=(t_L~b_L)^T$ and $t_R$ are the left-handed $SU(2)$ doublet of third generation and the right-handed $SU(2)$ singlet of top quark,
respectively, while $T_R$ and $T_L$ are two $SU(2)$ singlet top partners.

We obtain the mass matrix of top quark and the partner $(t,~T)$,
\begin{equation}
\left(\begin{array}{c} \bar{t}_L~~\bar{T}_L\\

\end{array}\right)\, \ \ \
\left(\begin{array}{c} m_t~~~~~0 \\
m'_T~~~m_T
\end{array}\right)
\left(\begin{array}{c} t_R \\
T_R
\end{array}\right).
\end{equation}
In this paper, we assume $m'_T$ to be very small so that there is no
mixing between $t$ and $T$. If the $\Phi_1$ has
the interactions with $Q_{tL}$ and $T_R$, the mixing of $t$ and $T$
will appear. Here we do not consider this case. Due to the absence of the mixing of $t$ and $T$,
the top partner mass is $m_T$. Using the Eq. (\ref{field}), the Eq. (\ref{partner}) gives the Yukawa
interactions,
\beq \label{weakcoupling}
\frac{y_T}{\sqrt{2}}\rho_2\bar{t}_L T_R - i \frac{y_T}{\sqrt{2}}A\bar{t}_L T_R - y_T H^- \bar{b}_L T_R + h.c..
\eeq
Using the Eq. (\ref{mixhiggs}), the Eq. (\ref{weakcoupling}) gives the top partner couplings to the Higgs bosons,
\bea\label{couplingt}
&&H\bar{t}_LT_R = H\bar{T}_Rt_L:~\frac{y_T}{\sqrt{2}}\sin\theta \nonumber\\
&&h\bar{t}_LT_R = h\bar{T}_Rt_L:~-\frac{y_T}{\sqrt{2}}\cos\theta \nonumber\\
&&A\bar{t}_LT_R = -A\bar{T}_Rt_L:~i\frac{y_T}{\sqrt{2}} \nonumber\\
&&H^-\bar{b}_LT_R = H^+\bar{T}_Rt_L:y_T. \eea
Due to the absence of the mixing of $t$ and $T$, there are not the diagonal couplings of $H\bar{T}T$ and $h\bar{T}T$.
For $\cos\theta \neq 0$, the $T\to t h$ channel will be open, and some
simulations on the channel at the LHC have been studied in \cite{tth}. In this paper we will take
$\cos\theta=0$ for which the coupling of $h\bar{t}T$ is absent.

For the singlet $T_L$ and $T_R$, the general neutral and charged current interactions are \cite{1205.4733}
\bea\label{wtbztt}
{\cal L}^{NC}=&&\frac{g}{c_W} Z_\mu \bar{t}~\left[(\frac{1}{2}-\frac{2}{3}s_W^2 +\delta g_L^t) P_L
+(-\frac{2}{3}s_W^2) P_R\right]~t\nonumber\\
&&+\frac{g}{c_W} Z_\mu \bar{T}~\left[(\frac{1}{2}-\frac{2}{3}s_W^2 +\delta g_L^T) P_L
+(-\frac{2}{3}s_W^2) P_R\right]~T\nonumber\\
&&+\frac{g}{c_W} Z_\mu \bar{T}~\left[\delta g_L^{Tt} P_L\right]~t +h.c.,\nonumber\\
{\cal L}^{CC}=&&\frac{g}{\sqrt{2}} W^{\mu+} c_{\theta_L}~\bar{t}~ \gamma_\mu P_L~b+
\frac{g}{\sqrt{2}} W^{\mu+} s_{\theta_L}~\bar{T}~ \gamma_\mu P_L~b+h.c..
\eea
Where $c_{\theta_L}=\cos\theta_L$ and $s_{\theta_L}=\sin\theta_L$ with $\theta_L$ being the mixing angle of
the left-handed top and the partner. $\delta g_L^t=-\frac{s_{\theta L}^2}{2}$,
$\delta g_L^T=-\frac{c^2_{\theta L}}{2}$ and $\delta g_L^{Tt}=\frac{s_{\theta_L}c_{\theta_L}}{2}.$
In this paper we assume that there is no mixing of $t$ and $T$, namely $s_{\theta_L}=0$. For this case, the Eq. (\ref{wtbztt}) shows
that the couplings of $Z\bar{T}t$ and $W^+ \bar{T}b$ are zero, and there are no decays of $T\to Zt$ and $T\to W^+ b$.

Similarly, we can introduce the bottom partner and the Yukawa interaction is given as
\beq - {\cal L} = y_B\, \bar{Q}_{tL} \, \Phi_2 B_R  + m_B\bar{B}_LB_R + m'_B\bar{B}_Lb_R
+ h.c.,
 \label{bpartner}\eeq
 where $B_R$ and $B_L$ are two $SU(2)$ singlet bottom partners. When $m'_B$ approaches to zero, there is no mixing between
 $b$ and $B$.
From the Eq. (\ref{bpartner}),  we can obtain the bottom partner couplings to Higgses,
\bea\label{couplingt}
&&H\bar{b}_LB_R = H\bar{B}_Rb_L:~\frac{y_B}{\sqrt{2}}\sin\theta \nonumber\\
&&h\bar{b}_LB_R = h\bar{B}_Rb_L:~-\frac{y_B}{\sqrt{2}}\cos\theta \nonumber\\
&&A\bar{b}_LB_R = -A\bar{B}_Rb_L:~-i\frac{y_B}{\sqrt{2}} \nonumber\\
&&H^+\bar{t}_LB_R = H^-\bar{B}_Rt_L:-y_B.
\eea
Similar to the top partner, there are no couplings of $h\bar{b}B$, $h\bar{B}B$, $H\bar{B}B$, $Z\bar{B}b$ and $W^-\bar{B}t$ for
$\cos\theta=0$ and the absence of the mixing of $b$ and $B$.

\section{Numerical calculations and discussions}
The 2HDM is usually described in the physical basis and Higgs basis. In the physical basis, the two Higgs doublet $\Phi_1$ and $\Phi_2$ have the
non-zero VEVs, and $\tan\beta$ is defined as $v_2/v_1$ with $v_1$ and $v_2$ being the VEV of the first and second scalar doublet. In the Higgs basis, the VEV of $\Phi_2$ is zero, therefore, the parameter $\tan\beta$
is absent. The coupling constants of Higgs potential in the Higgs basis as shown in the Eq. (\ref{V2HDM}), can be expressed
using the coupling constants and $\tan\beta$ in the physical basis \cite{0504050}.

In the Higgs basis, the Yukawa interactions of fermions are parameterized by the $\kappa_u$, $\kappa_d$ and $\kappa_\ell$ as shown in Eq. (\ref{yukawacoupling}) and
Eq. (\ref{normacoupling}), and which can be mapped to the four traditional types of 2HDMs via the $\kappa_u$, $\kappa_d$ and $\kappa_\ell$ specified in Table \ref{dltype}.
\begin{table}
 \caption{The specified values of $\kappa_u$, $\kappa_d$ and
$\kappa_\ell$ for the four traditional types of 2HDMs.}\vspace{0.5cm}
  \setlength{\tabcolsep}{2pt}
  \centering
  \begin{tabular}{|c|c|c|c|c|}
    \hline
     &~~Type~I~~& ~~~~Type~II~~~~ &~~Lepton-specific~~&~~~~ Flipped~~~~\\
    \hline
     ~~$\kappa_u$~~
     & $1/\tan\beta$ & $1/\tan\beta$
     & $1/\tan\beta$ & $1/\tan\beta$
     \\
     \hline
     ~~$\kappa_d$~~
      & $1/\tan\beta$ & $-\tan\beta$
      & $1/\tan\beta$ & $-\tan\beta$
     \\
     \hline
     ~~$\kappa_\ell$~~
      & $1/\tan\beta$ & $-\tan\beta$
      & $-\tan\beta$ & $1/\tan\beta$
     \\
     \hline
      \end{tabular}
\label{dltype}
\end{table}
However, here we take $\sin\theta=1$, $\kappa_d=\kappa_\ell$=0.
For this choice, the Eq. (\ref{normacoupling}) shows that
the 750 GeV Higgs ($H$) couplings to the down-type quark and lepton are zero, and the coupling to gauge boson is zero, which
can naturally satisfy the bounds from the measurements of the diboson, dijet and dilepton.
The 125 GeV Higgs ($h$) couplings to up-type quark, down-type quark, lepton and gauge boson are the
same as the SM Higgs, and the couplings to the
new charged Higgs, $T$ and $B$ quark equal to zero for $\sin\theta=1$.
The $H$ coupling to the up-type quark is proportional to $\kappa_u$, which can control
the width of $H\to \bar{t}t$ and is taken as a free input parameter. However, the experimental data
 of the 750 GeV Higgs diphoton rate will give the upper bound of the width of the 750 GeV Higgs.

\subsection{$T$ and $B$ decay}
As discussed above, the partners $T$ and $B$ have no couplings to the
gauge bosons and the 125 GeV Higgs in the parameter space taken in this
paper. Therefore, $T$ and $B$ can be hardly constrained by the
current experimental data of the exotic quark from the ATLAS
\cite{atlast} and CMS \cite{cmst} searches. The main decay modes are
$T\to t H$, $T\to t A$ and $T\to b H^+$ for the $T$ quark, as well
as $B\to b H$, $B\to b A$ and $B\to t H^-$ for the $B$ quark. For
$m_H=750$ GeV, the oblique parameters favor $m_{H^{\pm}}$ and $m_A$
to have the degenerate mass, especially for that their mass have
sizable deviation from 750 GeV.

\begin{figure}[tb]
 \epsfig{file=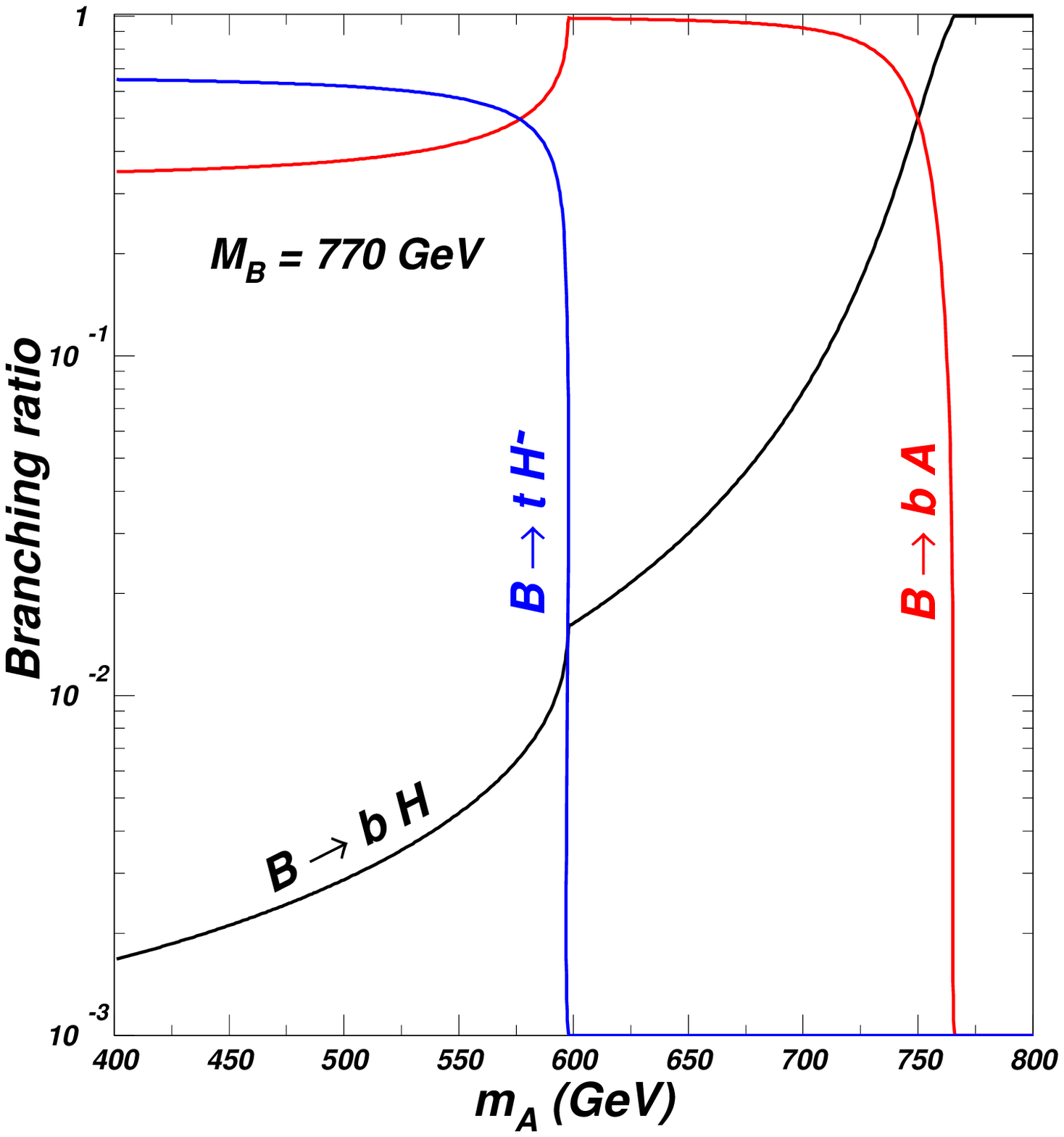,height=7.5cm}
 \epsfig{file=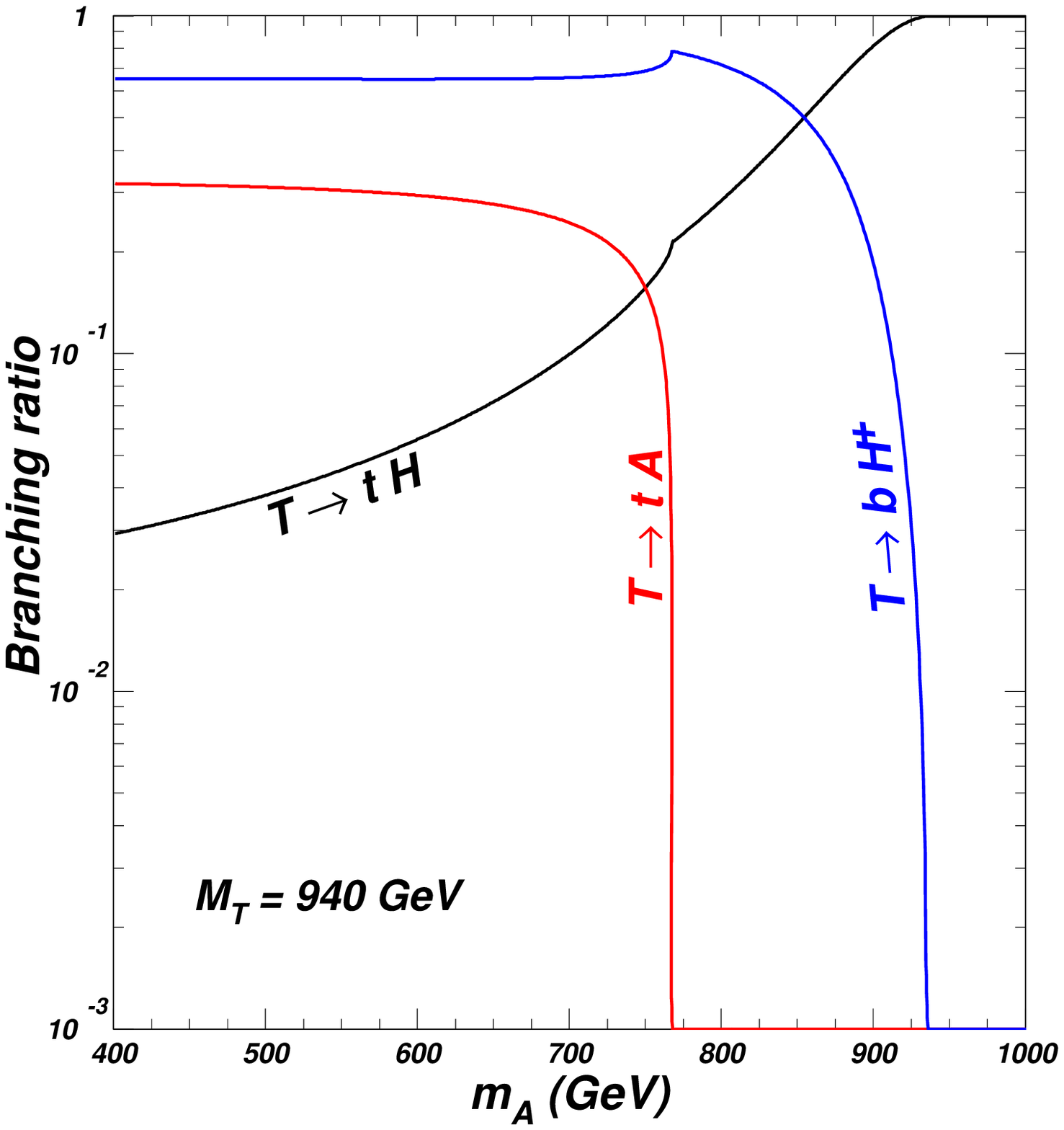,height=7.5cm}
\vspace{-0.3cm} \caption{The branching ratios of $T$ and $B$ versus $m_A$ with $m_{H^{\pm}}=m_A$ and
 $m_H=750$ GeV.} \label{br}
\end{figure}

We take $m_B=$ 770 GeV and $m_T=$ 940 GeV, and plot their branching ratios versus $m_A$ in Fig. \ref{br}.
Since the widths of $T\to t H$, $T\to t A$ and $T\to b H^+$ are proportional to $y_T^2$, their branching
ratios are independent on $y_T$, which also holds for the $B$ quark and $y_B$. Both $Br(B\to bH)$ and $Br(T\to t H)$ are
very small for $m_A$ and $m_{H^{\pm}}$ are much smaller than $m_H$, and increase with
$m_A$ and $m_{H^{\pm}}$. For $m_B=$ 770 GeV and $m_H=m_A=m_{H^{\pm}}=$ 750 GeV, $B\to tH^{-}$ is kinematically forbidden, and
$Br(B\to bH)$ and $Br(B\to bA)$ have the same value and equal to 50\% nearly.
For $m_T=$ 940 GeV and $m_H=m_A=m_{H^{\pm}}=$ 750 GeV, $T\to b H^{+}$ dominates over $T\to t H$
and $T\to t A$ since the former has an enhanced factor of 2 from the coupling, and $T\to t H$
and $T\to t A$ are suppressed by a large phase space.
Only for $m_{H^{\pm}}$ and $m_A$ are very closed to $m_T$ and $m_B$, $T\to t H$ and $B\to b H$ are the dominant decay modes.

\subsection{The production rate of 750 GeV diphoton resonance}
\begin{figure}[tb]
 \epsfig{file=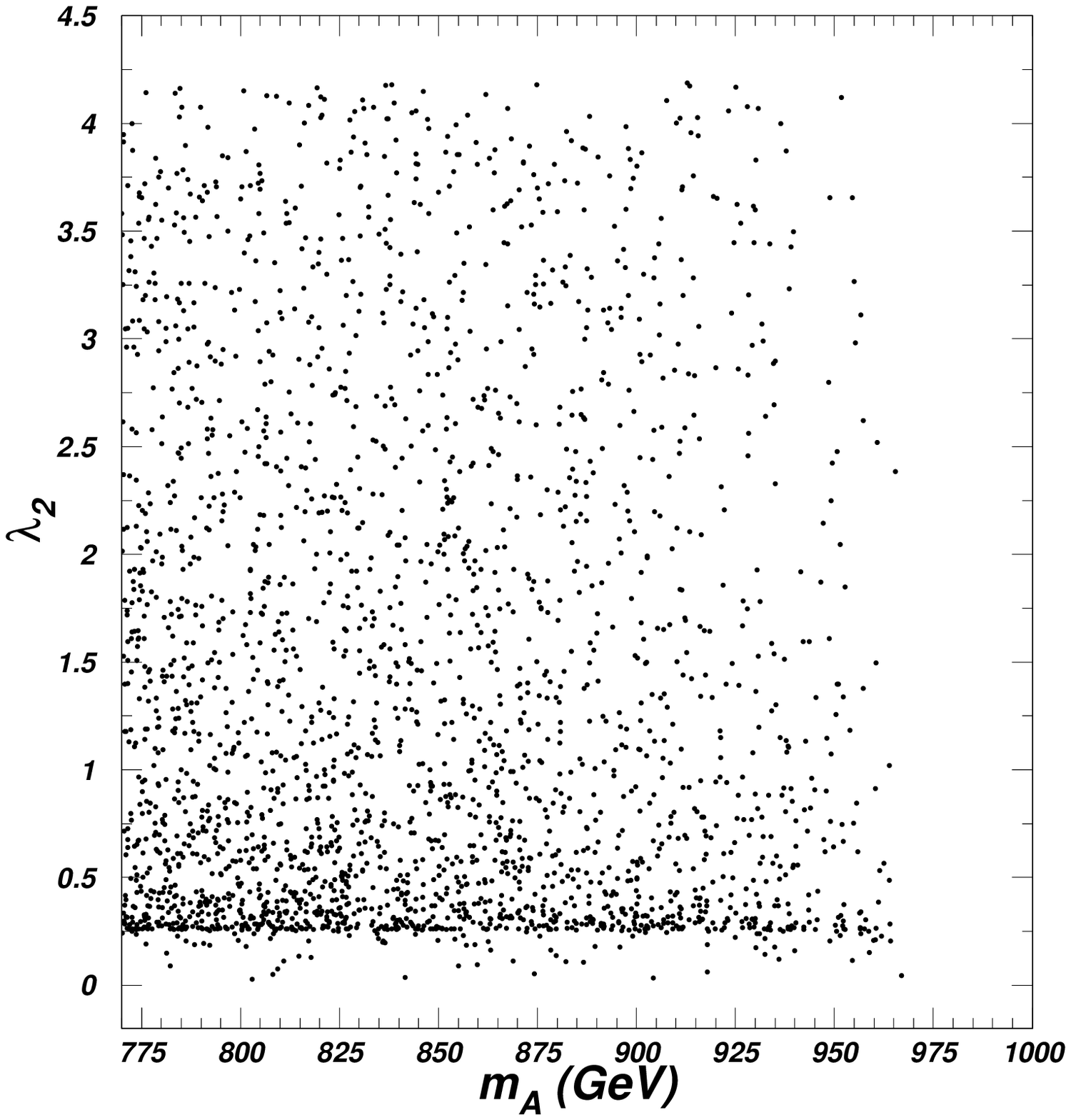,height=5.65cm}
 \epsfig{file=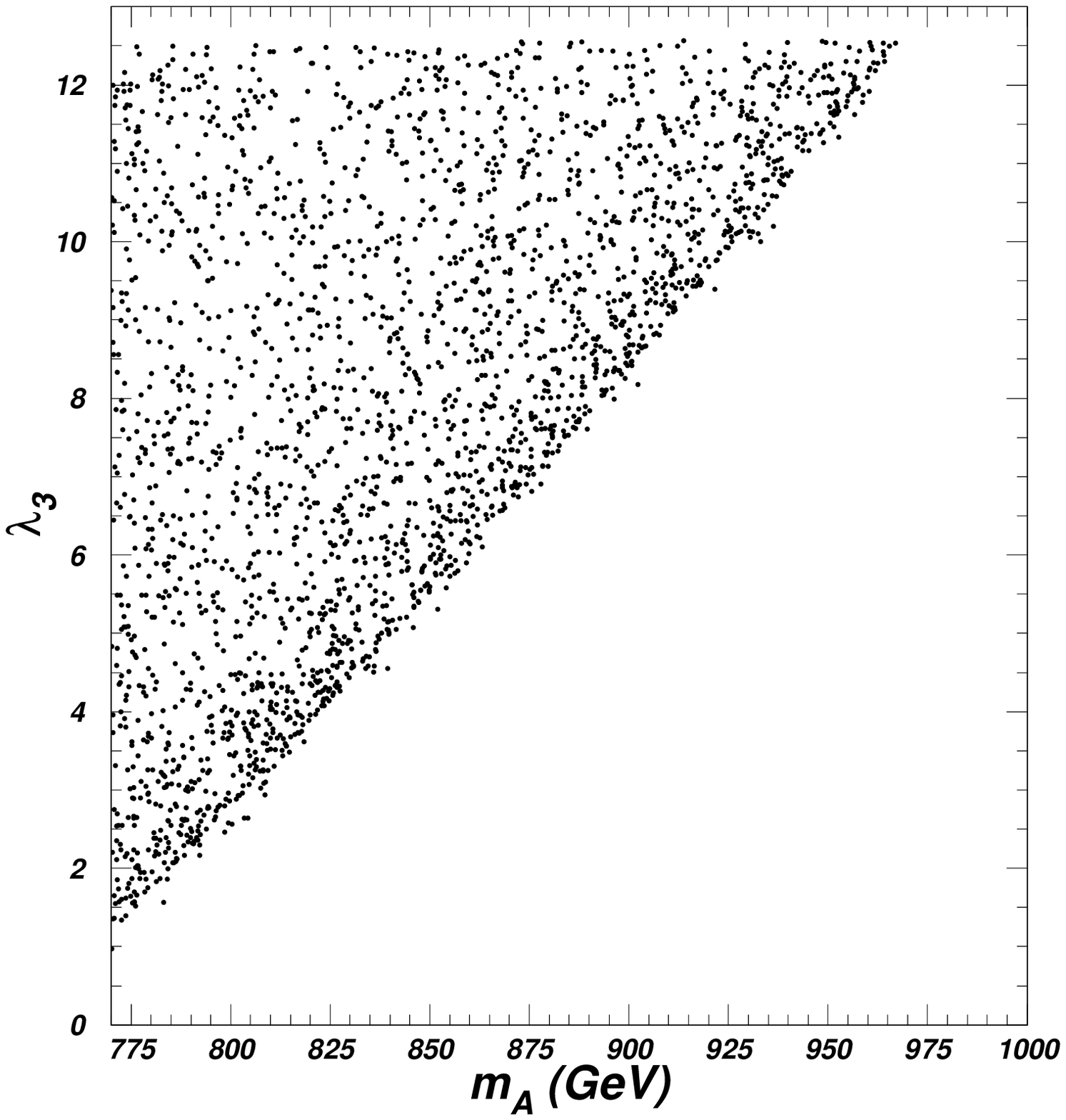,height=5.65cm}
 \epsfig{file=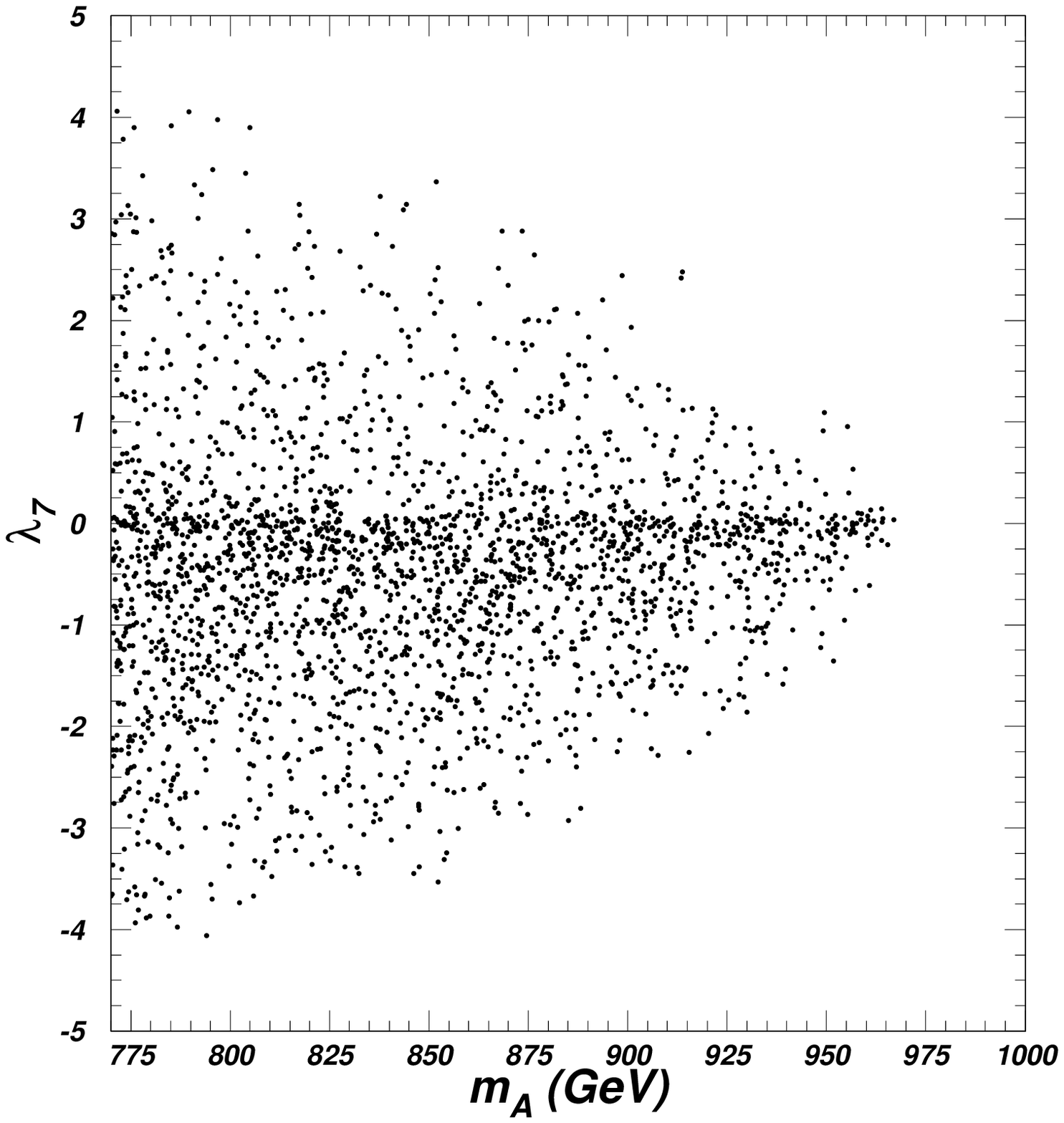,height=5.65cm}
\vspace{-1.0cm} \caption{The samples allowed by the theoretical constraints from the vacuum stability, unitarity and perturbativity
for $m_A=m_{H^{\pm}}$, $m_H=750$ GeV and $\cos\theta=$0.} \label{theo}
\end{figure}

In order to obtain the maximal production rate, we assume
$m_A=m_{H^{\pm}}$ to be larger than $m_B$ and $m_T$, which leads
$Br(B\to b H)$=$Br(T\to t H)$=1. Also $H\to AZ$, $H\to H^\pm W^\mp$,
$H\to AA$ and $H\to H^+H^-$ are kinematically forbidden for this
case. For $m_A=m_{H^{\pm}}$, $\lambda_4$ and $\lambda_5$ are
determined by $m_H$ and $m_{H^{\pm}}$ from the Eq. (\ref{mamhp}) and
Eq. (\ref{mhmhp}), \beq
\frac{\lambda_4}{2}=\lambda_5=\frac{m_H^2-m_{H^{\pm}}^2}{2v^2}. \eeq
For $m_{H^{\pm}}>m_H$, $\lambda_4$ and $\lambda_5$ are negative,
which will be constrained by the vacuum stability to some extent.
As discussed in the Section II, $\cos\theta=0$ determines $\lambda_6=0$ and
$\lambda_1=\frac{m_h^2}{2v^2},$ and
 $\lambda_2$, $\lambda_3$ and $\lambda_7$ are the free parameters, which can be tuned to satisfy
theoretical constraints from the vacuum stability, unitarity and perturbativity.
Refs. \cite{stab1,stab2,unit1,unit2} give the corresponding well-known classical formulas for the constrains on the coupling constants of
the physical basis. We employ $\textsf{2HDMC}$ \cite{2hc-1}
to perform the theoretical constraints on the coupling constants in the physical basis, and then
use the formulas of Eqs. (A16-A22) in the ref. \cite{0504050} to transform the results into the constraints on $\lambda_2$, $\lambda_3$ and $\lambda_7$
 in the Higgs basis, namely expressing $\lambda_2$, $\lambda_3$ and $\lambda_7$ with the allowed parameters of the physical basis. In the Fig. \ref{theo},
 we project the samples allowed by the theoretical constraints
 on the planes of $\lambda_2$ versus
$m_A$, $\lambda_3$ versus $m_A$ and $\lambda_7$ versus $m_A$ for $m_A=m_{H^{\pm}}$, $m_H=750$ GeV and $\cos\theta=$0. Fig. \ref{theo}
shows that $\lambda_2$ is required to be larger than 0. With the increasing of $m_A$ and $m_{H^{\pm}}$, the absolute values of $\lambda_4$
and $\lambda_5$ become large ($\lambda_4$ and $\lambda_5$ are negative), which favors the large $\lambda_3$ and the $\lambda_7$ with a small
absolute value.

\begin{figure}[tb]
 \epsfig{file=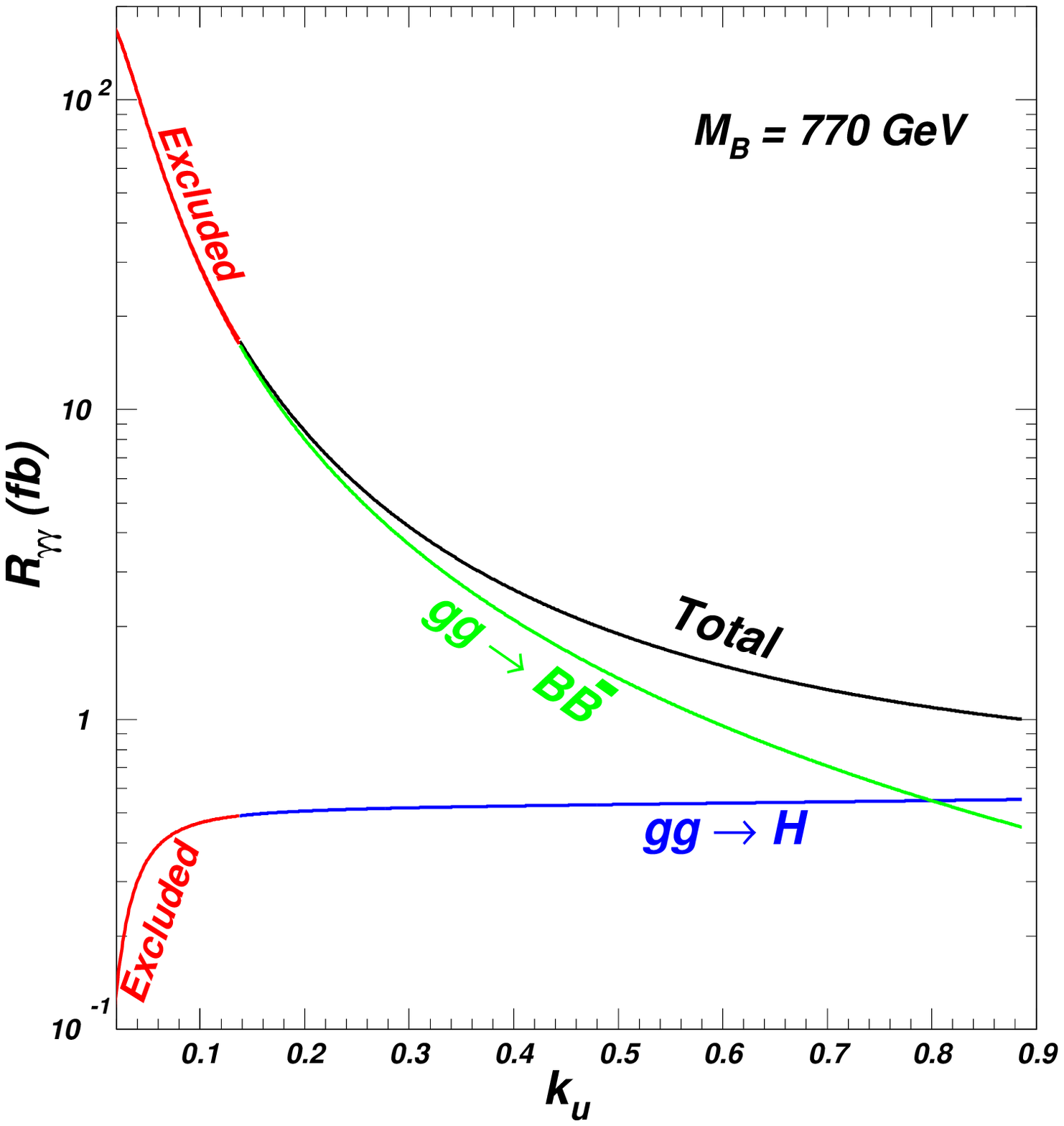,height=7.5cm}
 \epsfig{file=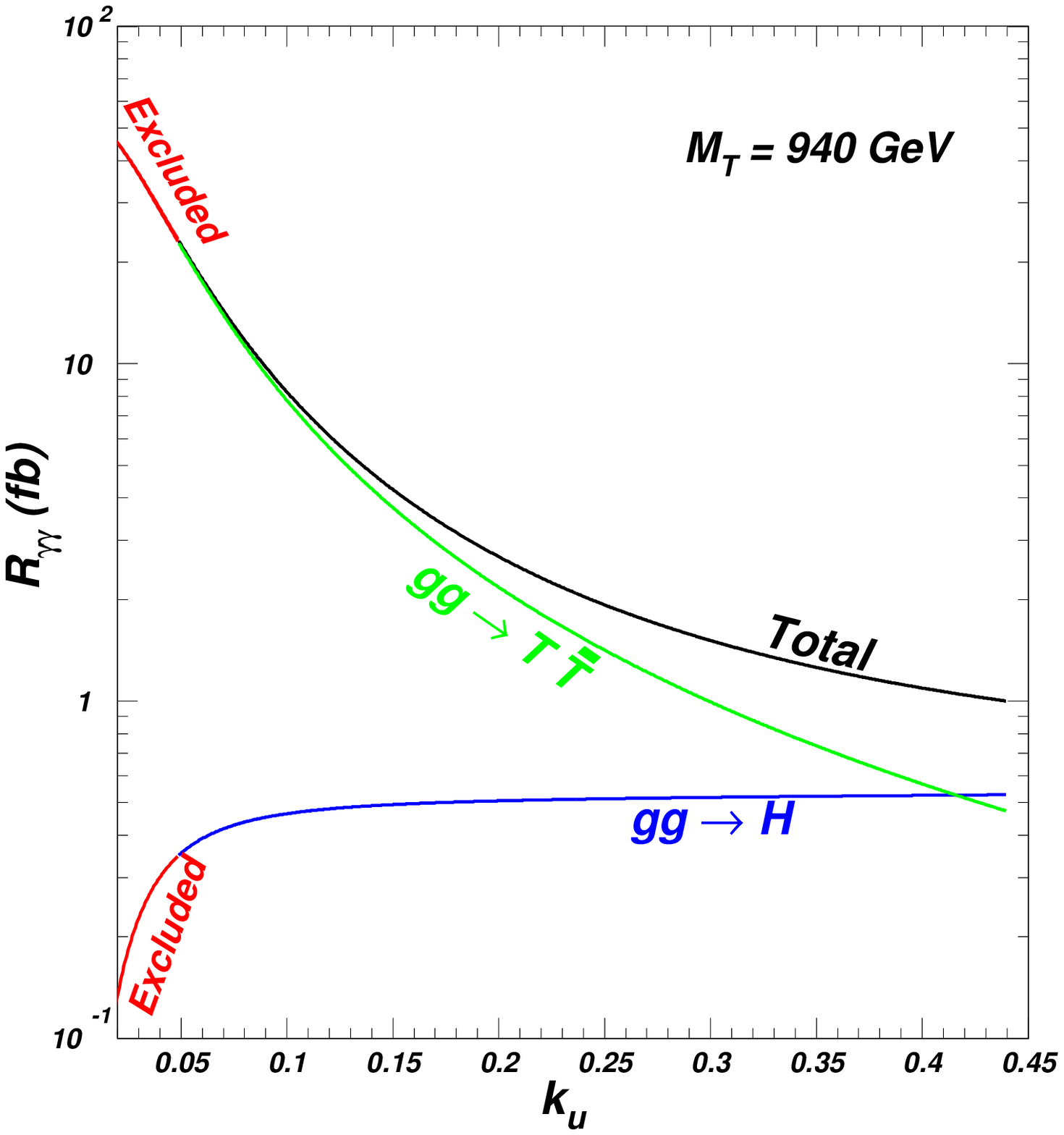,height=7.65cm}
\vspace{-0.3cm} \caption{The diphoton production rate of the 750 GeV Higgs versus $k_u$ for $m_B=$ 770 GeV and
$m_T=$ 940 GeV. The 750 GeV Higgs coupling to the new charged Higgs ($k_4$) is fixed as $4\pi$, and $m_\pi=m_\chi =$375 GeV.} \label{sigallb}
\end{figure}

The widths of $H\to WW,~ZZ,~hh,~b\bar{b},~\tau\bar{\tau}$ at the tree-level are
zero for $\cos\theta=$ 0 and $\kappa_d=\kappa_\ell=0$.
 Therefore, $H\to t\bar{t}$ is the dominant decay
mode for a large $k_u$. Also the one-loop decays $H\to gg$, $H\to
\gamma\gamma$, $H\to Z\gamma$ and $H\to ZZ$ are considered, and the
last three modes can be sizably enhanced by the new charged scalars
at the one-loop. Since the $\pi^+$ and $\chi^{++}$ are $SU(2)_L$
singlets, for the charged scalars give the dominant contributions to
$H\to \gamma\gamma$, there is an approximate relation, \beq
\Gamma(H\to \gamma\gamma) :\Gamma(H\to Z\gamma):\Gamma(H\to ZZ)
=1:0.6:0.09. \eeq

We define the production rate of the 750 GeV diphoton, \bea
R_{\gamma\gamma}&\equiv&\sigma(gg\to B\bar{B}~(T\bar{T}))\times
Br(B\bar{B}~(T\bar{T})\to HHb\bar{b}~(t\bar{t}))\times Br(HH\to \gamma\gamma+X)\nonumber\\
&&+\sigma(gg\to H)\times
Br(H\to \gamma\gamma)\nonumber\\
&=&\sigma(gg\to B\bar{B}~(T\bar{T}))\times
 Br(HH\to \gamma\gamma+X)+\sigma(gg\to H)\times
Br(H\to \gamma\gamma).
\eea

At the LHC, the cross sections of $gg\to B\bar{B}~(T\bar{T})$ with $m_B=$ 770 GeV ($m_T=$ 940 GeV) are approximate
240 (65) fb for $\sqrt{s}$=13 TeV and 28 (5.5) fb for $\sqrt{s}$=8 TeV \cite{1512.04536}.

In our calculations, we consider the relevant collider bounds from
LHC searches at $\sqrt{s}$=8 TeV \cite{tt,rr,zr,zz}:
\bea &&\sigma_{t\bar{t}} < 550~ {\rm fb},~~~~~ \sigma_{\gamma\gamma} < 2~ {\rm fb},\nonumber\\
&&\sigma_{Z\gamma} < 4~ {\rm fb},~~~~~~ \sigma_{ZZ} < 12~ {\rm fb}.
\eea

Taking $k_4=4\pi$, $m_\pi=m_\chi=375$ GeV, $m_T$ = 940 GeV and $m_B$ = 770 GeV, we project the surviving samples on the plane
 of $R_{\gamma\gamma}$ versus $\kappa_u$ in Fig. \ref{sigallb}. Since the heavy CP-even Higgs coupling to top
quark is proportional to $\kappa_u$, the production rate from $gg\to
H$ increases with $\kappa_u$. Since the cross section of $gg\to
B\bar{B}~(T\bar{T})$ is
 independent on $\kappa_u$, and the total width of 750 GeV Higgs increases with $\kappa_u$, the production rate from
 $gg\to B\bar{B}~(T\bar{T})$ decreases with increasing of $\kappa_u$. The production rate from latter dominates over
 the former for the small $\kappa_u$, and equals to the former for $\kappa_u=$ 0.8 (0.42).
$R_{\gamma\gamma}>$ 1 fb favors $\kappa_u$ to be smaller than 0.9 for $m_B=$ 770 GeV and 0.45 for $m_T=$ 940 GeV.
 Compared to the bottom partner, the top partner mass is required to be larger than 930 GeV to open the decay $T\to t H$.
 The cross section of $gg\to T\bar{T}$ with $m_T=$ 940 GeV is 65 fb at the LHC with $\sqrt{s}=$ 13 TeV, which is much smaller than that
of $gg\to B\bar{B}$ with $m_B=$ 770 GeV, 240 fb. Therefore, the constraints on $m_T=$ 940 GeV are more strong than those on $m_B=$ 770 GeV.

For a very small top Yukawa coupling, the total width of 750 GeV
Higgs is very narrow, which leads a large $Br(H\to \gamma\gamma)$.
Therefore, the very small $\kappa_u$ is mainly excluded by the
experimental data of the diphoton rate at the $\sqrt{s}=$ 8 TeV, and
the lower bound of $\kappa_u$ is 0.14 for $m_B$= 770 GeV and 0.05
for $m_T=$ 940 GeV.

\begin{figure}[tb]
 \epsfig{file=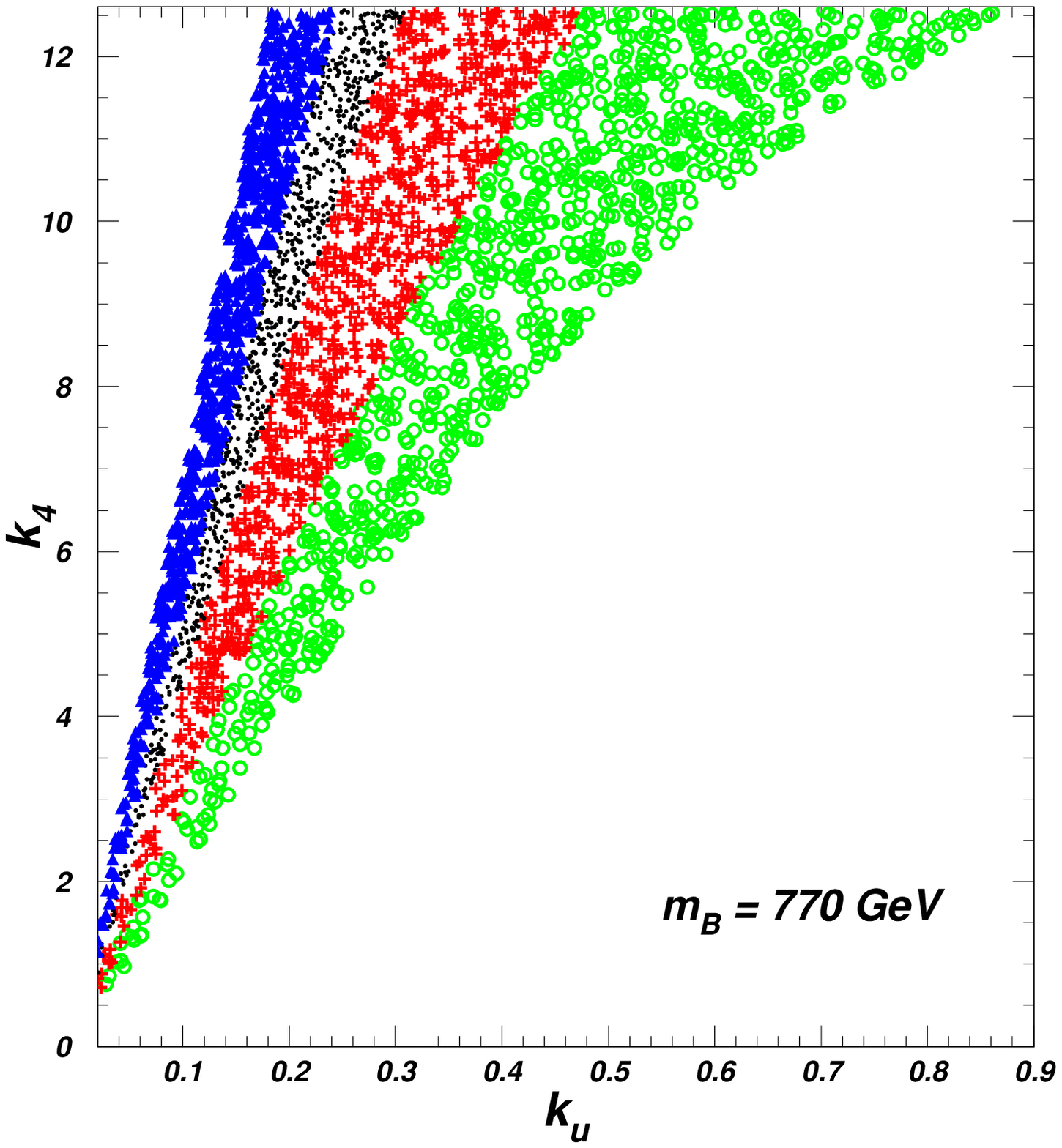,height=5.7cm}
  \epsfig{file=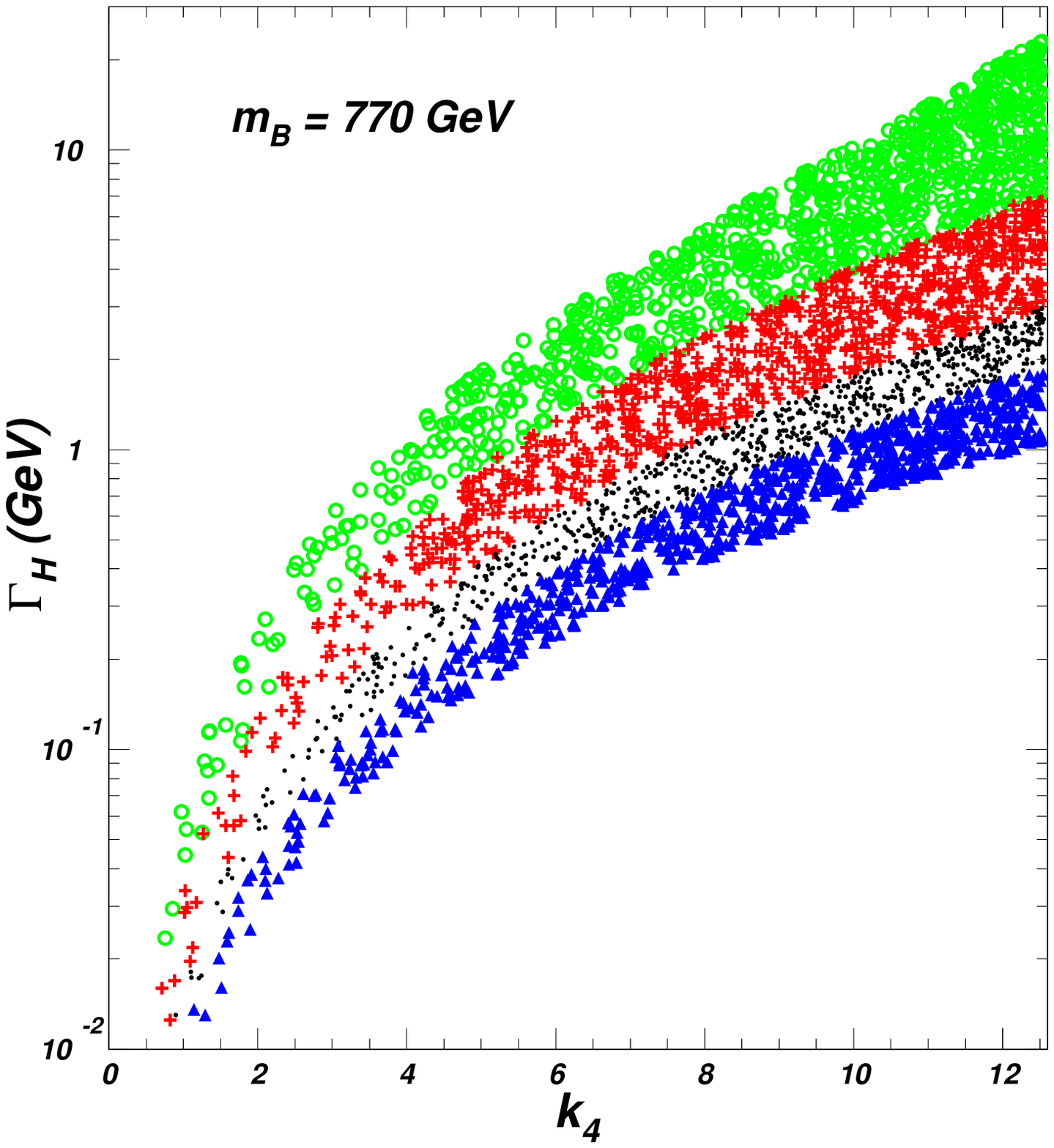,height=5.7cm}
 \epsfig{file=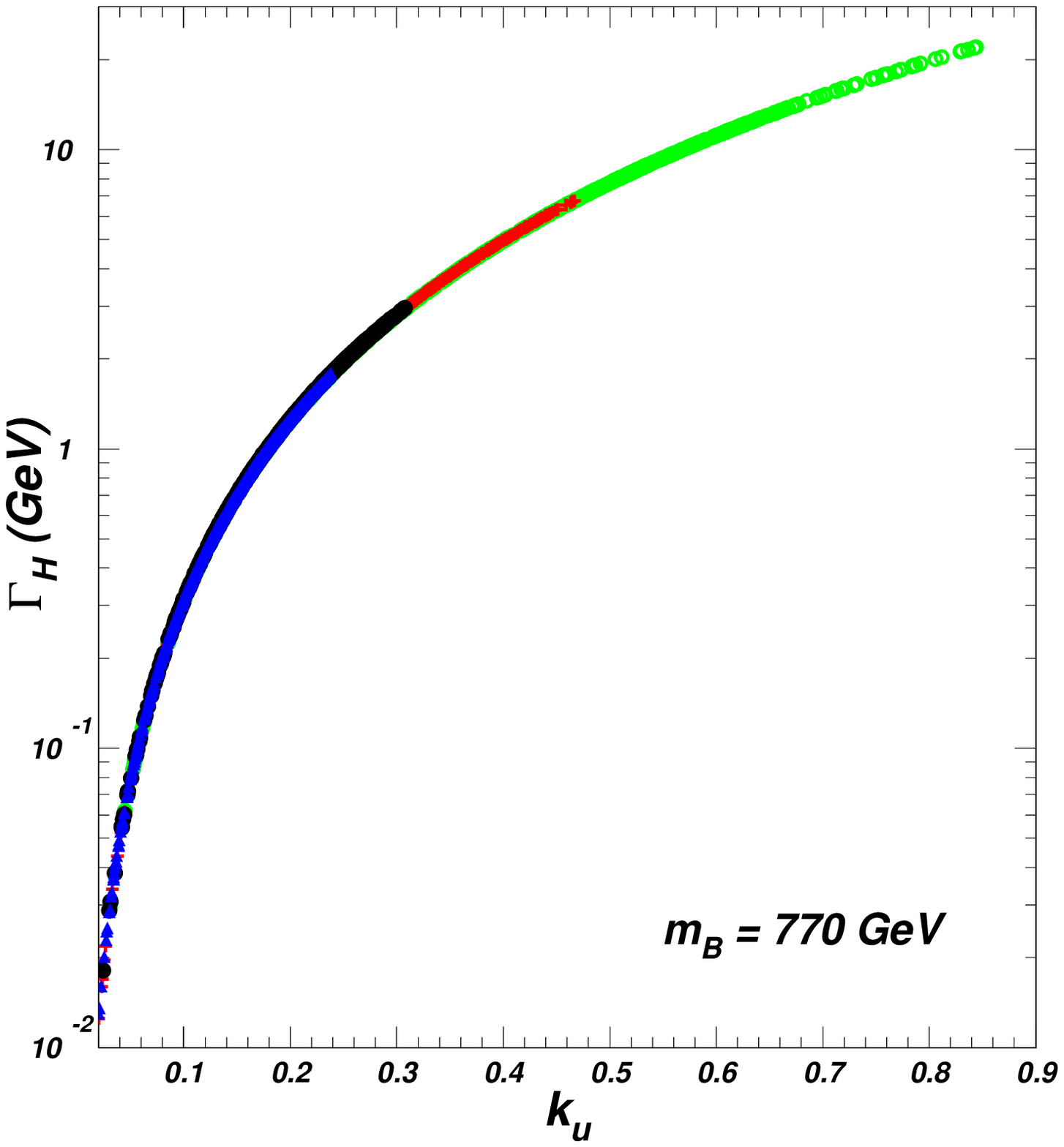,height=5.7cm}
 \epsfig{file=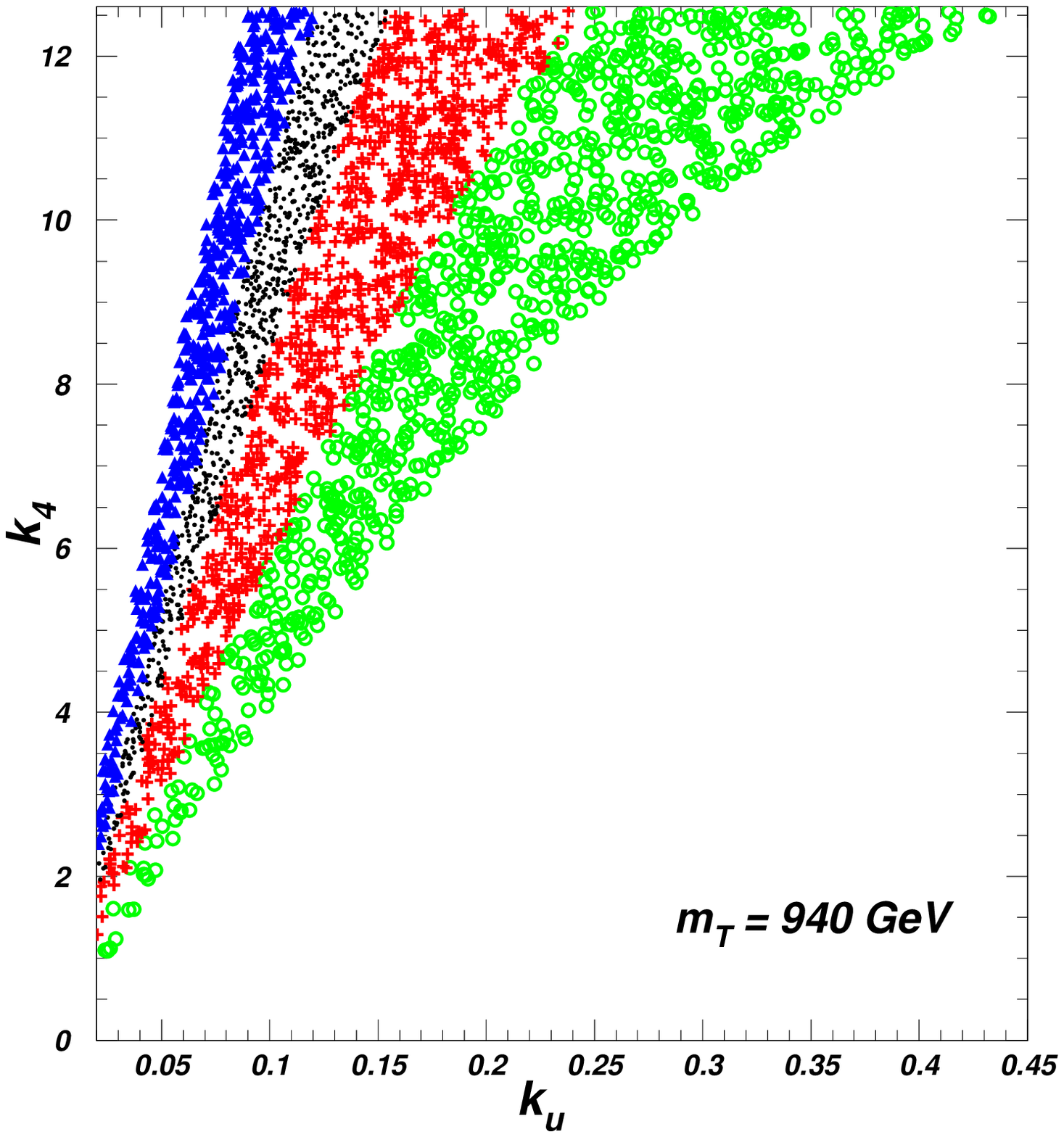,height=5.7cm}
  \epsfig{file=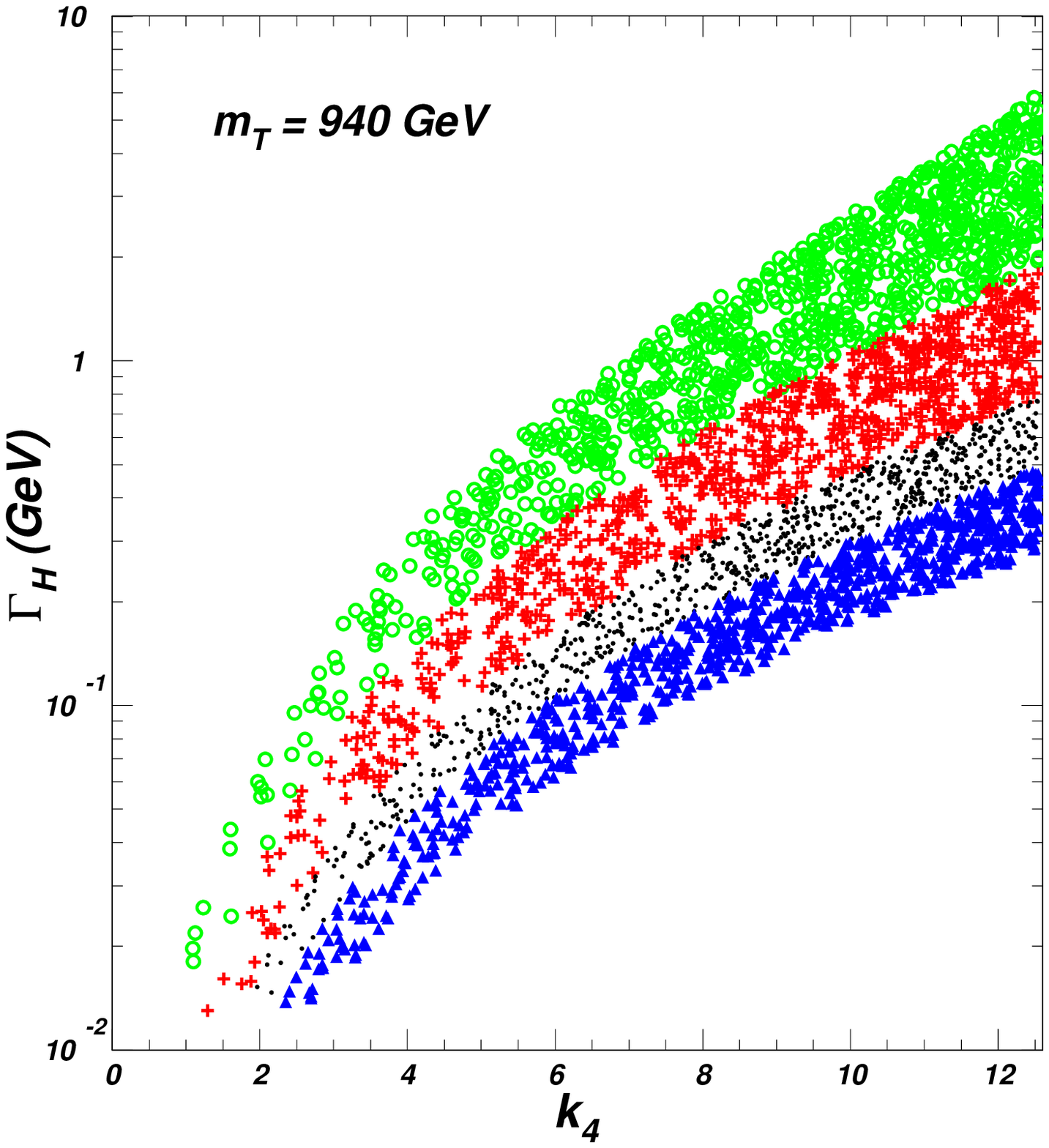,height=5.7cm}
 \epsfig{file=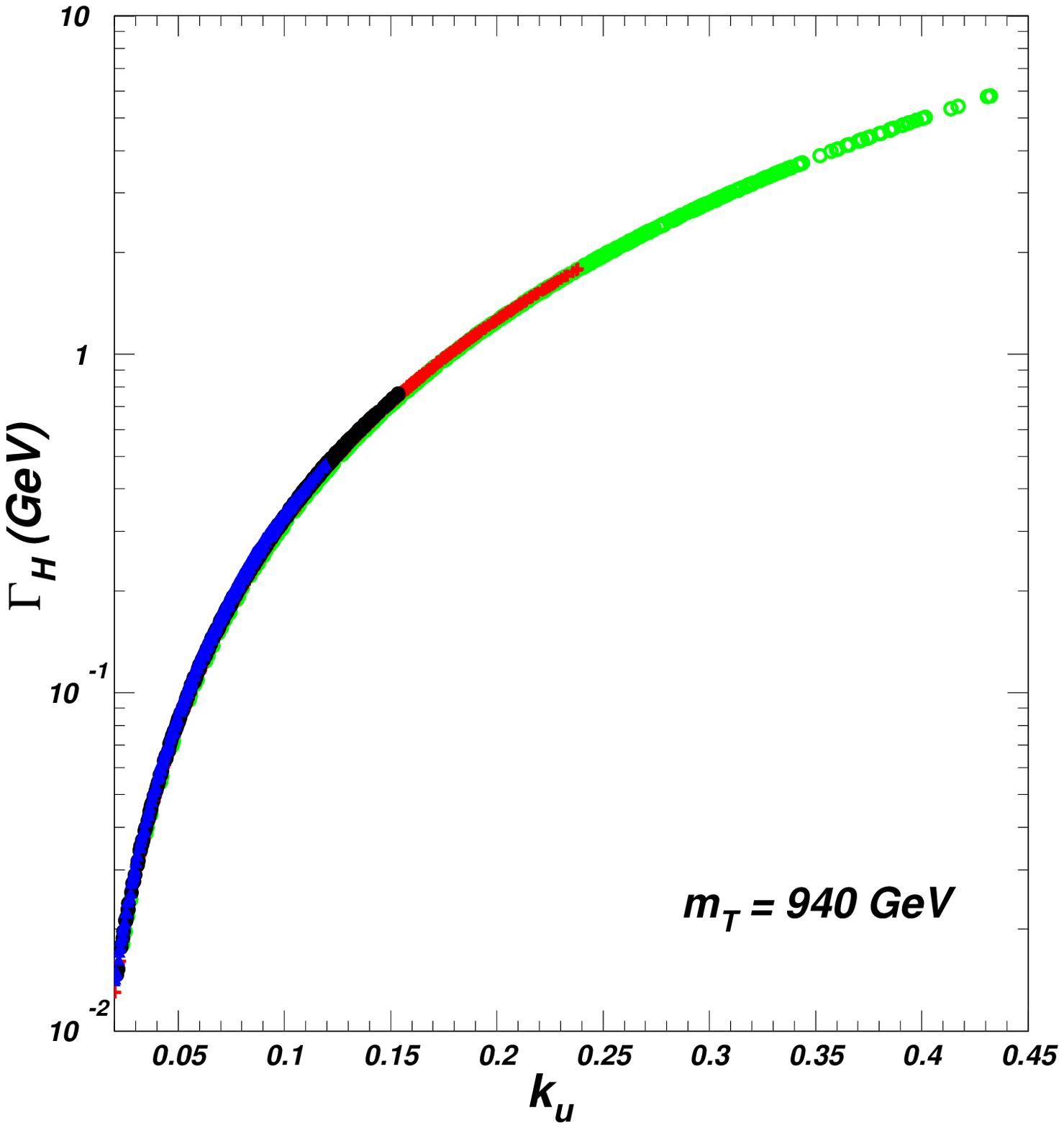,height=5.7cm}
\vspace{-0.5cm} \caption{The surviving samples projected on the
planes of $\kappa_4$ versus $\kappa_u$, $\Gamma_H$ versus $\kappa_4$
and $\Gamma_H$ versus $\kappa_u$ for $m_B=$ 770 GeV and $m_T=$ 940
GeV, with 1 fb $<R_{\gamma\gamma}<$ 2 fb for the circles (green), 2 fb
$<R_{\gamma\gamma}<$ 4 fb for the pluses (red), 4 fb
$<R_{\gamma\gamma}<$ 6 fb for the bullets (black) and 6 fb
$<R_{\gamma\gamma}<$ 10 fb for the triangles (blue).} \label{mb}
\end{figure}

In Fig. \ref{mb}, we project the surviving samples on the planes of
$\kappa_4$ versus $\kappa_u$, $\Gamma_H$ versus $\kappa_4$ and
$\Gamma_H$ versus $\kappa_u$. The large $R_{\gamma\gamma}$ favors a
small $\kappa_u$ and a large $k_4$, and the former can suppress the
total width of the 750 GeV Higgs, and the latter can enhance the
decay $H\to \gamma\gamma$ via the charged Higgs couplings to the 750
GeV Higgs. $R_{\gamma\gamma} >$ 2 fb favors $\kappa_u <$ 0.46 and
$k_4>$ 0.05 for $m_B=$ 770 GeV, and $\kappa_u <$ 0.24 and $k_4>$ 1.0
for $m_T=$ 940 GeV. For $m_B=$ 770 GeV, the total width can reach 7
GeV for $R_{\gamma\gamma} =$ 2 fb, and be larger than 2 GeV for
$R_{\gamma\gamma} <$ 6 fb. For $m_T=$ 940 GeV, the total width can
reach 2 GeV for $R_{\gamma\gamma} =$ 2 fb, and be larger than 0.8
GeV for $R_{\gamma\gamma} <$ 4 fb.

With the increasing of the mass of bottom partner and top partner, the cross section of $gg\to B\bar{B}~(T\bar{T})$ will
decrease rapidly and be around 2 fb for $m_B=m_T=$ 1500 GeV. For the enough small top quark Yakawa coupling of
the 750 GeV Higgs, the 750 GeV Higgs will mainly decay into  $\gamma\gamma$, $Z\gamma$ and $ZZ$, and $Br(H\to \gamma\gamma)$
is around $60\%$. Therefore, the diphoton production rate of the 750 GeV Higgs will reach 1.2 fb for $m_B=m_T=$ 1500 GeV.
For $m_B$ and $m_T$ are smaller than 1500 GeV, the more large production rate can be obtained. However, the
total width of the 750 GeV Higgs is required to be very narrow to enhance the production rate.
 Therefore, the precise measurement of width at the LHC can be as
a sensitive probe of the bottom partner and top partner.

For the 750 GeV resonance, the CMS slightly prefers a narrow width, and the ATLAS favors a width of 45 GeV.
Such large width can be obtained by the enhancement of $\kappa_u$ and $\kappa_d$ which can enhance the
widths of $H\to \bar{t}t,~\bar{b}b$. However, with
the increasing of the total width, the branching ratio of the diphoton mode decreases, which will suppress the
diphoton production rate. For the model with one singlet bottom partner ($m_B=770$ GeV), the total width of the 750 GeV Higgs is required
to be smaller than 7 GeV in order to obtain $R_{\gamma\gamma}>$ 2 fb. Some additional charged particles need be introduced to enhance the
width of $H\to \gamma\gamma$ in order to obtain $R_{\gamma\gamma}>$ 2 fb and $\Gamma_H\simeq$ 45 GeV.

In this paper, we discussed the two different scenarios of the singlet top partner and the singlet bottom partner.
Besides, one can attempt to introduce the doublet fields,
\begin{equation} \label{field2}
\Psi'_L=\left(\begin{array}{c} T_L \\
B_L
\end{array}\right)\,, \ \ \
\Psi'_R=\left(\begin{array}{c} T_R \\
B_R
\end{array}\right).
\end{equation}
The Yukawa interactions can be given as
\beq - {\cal L} = y_T\, \bar{\Psi}'_{L} \, \widetilde\Phi_2 t_R + y_B\, \bar{\Psi}'_{L} \, \Phi_2 b_R
+ m \bar{\Psi}'_{L} \Psi'_{R}+ m'\bar{Q}_{tL}\Psi'_{R}+h.c..
 \label{bpartner2}\eeq
In order to avoid the experimental constraints of the ATLAS and CMS
searches for the $T\to Wb$, $T\to tZ$, $T\to th$, $B\to Wt$, $B\to
bZ$ and $B\to bh$, one can assume that there are no mixings of $t$
and $T$ as well as $b$ and $B$, namely $m'=0$. For this case, the
$T$ and $B$ have the degenerate mass, \beq m_T=m_B=m, \eeq and the
charged current of $T$ and $B$ still appears since they are the
doublets of $SU(2)_L$, \beq {\cal L}^{CC}=\frac{g}{\sqrt{2}}
W^{\mu+} \bar{T}~ \gamma_\mu ~B+h.c.. \eeq In order to obtain the
large cross sections of $pp\to \bar{B}B/\bar{T}T$, one should take
the small masses of $B$ and $T$. For $m_T=m_B=770$ GeV, the 750 GeV
diphoton production rate from $pp\to \bar{B}B\to HH\bar{b}b$ process
is the same as the model with the singlet $B_L$ and $B_R$. Also the
750 GeV diphoton resonance can be originating from the $pp\to \bar{T}T$
followed by the off-shell decays $T\to t^* H$ and $T\to W^* B^* (\to
Hb)$. Since $T$ will partly decay into the other objects, including
the off-shell 750 GeV Higgs, the 750 GeV diphoton rate from the
$pp\to \bar{T}T$ is smaller than that of the $pp\to \bar{B}B$. This
model predicts the existence of the top partner and bottom partner
simultaneously, and the particle spectrum is more complicated than
the model with one top partner and the model with one bottom partner
which are studied in this paper.

\section{Conclusion}
To accommodate the 750 GeV diphoton excess, we proposed an extension
of 2HDM with the top and bottom partners. In addition, we took the
approach of Zee-Babu model to introduce two scalar singlets (one
is singly charged, and the other is doubly charged), which can naturally
give a small neutrino Majorana mass and enhance the 750 GeV Higgs decay into diphoton.
In this model, the
production rate $R_{\gamma\gamma}$ of the 750 GeV diphoton is from
both $gg\to B\bar{B}~(T\bar{T})$ and $gg\to H$, and the former
dominates over the latter for a small top quark coupling with the
750 GeV Higgs, and is comparable to the latter for a large top
Yukawa coupling.

For $m_B=$ 770 GeV, $R_{\gamma\gamma} >$ 2 fb favors $\kappa_u <$ 0.46 and
$k_4>$ 0.05, and the total width of the 750 GeV Higgs can reach 7 GeV for $R_{\gamma\gamma} =$ 2 fb.
For $m_T=$ 940 GeV, $R_{\gamma\gamma} >$ 2 fb favors $\kappa_u <$ 0.24 and
$k_4>$ 1.0, and the total width can reach 2 GeV for $R_{\gamma\gamma} =$ 2 fb.
To obtain enough large production rate of the 750 GeV diphoton, the total width tends
to decrease with the increasing of the bottom partner and top partner masses. Therefore, the precise
measurement of the width of the resonance at the LHC can be as
a sensitive probe of these bottom partner and top partner.

\section*{Acknowledgment}
This work has been supported in
part by the National
Natural Science Foundation of China under grant Nos. 11575152, 11305049, 11275057,
11405047, 11275245, 10821504 and 11135003,  by
Specialized Research Fund for the Doctoral Program of
Higher Education under Grant No.20134104120002, and by the Spanish
Government and ERDF funds from the EU Commission
[Grants No. FPA2014-53631-C2-1-P, SEV-2014-0398, FPA2011-23778], and
by the Australian Research Council, by the CAS Center for Excellence in Particle Physics
(CCEPP).

\end{document}